\providecommand{\PaperRoot}{.}
\documentclass[twocolumn]{\PaperRoot/aastex701} 
\graphicspath{{\PaperRoot/Figs/}}

\usepackage{amsmath}
\usepackage{CJK}

\definecolor{HHcolor1}{rgb}{0.93,0.57,0.13}
\definecolor{HHcolor2}{rgb}{0.5,0.1,0.5}
\definecolor{HHcolor3}{rgb}{0.0,0.0,1.0}

\usepackage[dvipsnames]{xcolor}

\begin{document}
\begin{CJK*}{UTF8}{gbsn}

\title{Clumpy Outflows from Super-Eddington Accreting Black Holes. II. Conditions for Clump Formation}

\author[orcid=0000-0003-3143-3995, gname='Haojie', sname='Hu']{Haojie Hu (胡豪杰)}
\affiliation{Center for Computational Science, University of Tsukuba, 1-1-1 Tennodai, Tsukuba, Ibaraki 305-8577, Japan}
\email[show]{huhaojie@ccs.tsukuba.ac.jp}  

\author[orcid=0000-0003-3640-1749, gname='Yuta', sname='Asahina']{Yuta Asahina} 
\affiliation{Center for Computational Science, University of Tsukuba, 1-1-1 Tennodai, Tsukuba, Ibaraki 305-8577, Japan}
\affiliation{Department of Natural Science, Faculty of Arts and Science, Komazawa University, 1-23-1 Komazawa, Setagaya-Ku, Tokyo 154-8525, Japan}
\email{asahinyt@ccs.tsukuba.ac.jp}

\author[orcid=0009-0005-6822-4179, gname='Shogo', sname='Yoshioka']{Shogo Yoshioka} 
\affiliation{Center for Computational Science, University of Tsukuba, 1-1-1 Tennodai, Tsukuba, Ibaraki 305-8577, Japan}
\affiliation{Department of Astronomy, Graduate School of Science, Kyoto University, Kitashirakawa-Oiwake-cho, Sakyo-ku, Kyoto 606-8502, Japan}
\email{yoshioka@kusastro.kyoto-u.ac.jp}

\author[orcid=0000-0003-0114-5378, gname=Hiroyuki R., sname='Takahashi']{Hiroyuki R. Takahashi} 
\affiliation{Department of Natural Science, Faculty of Arts and Science, Komazawa University, 1-23-1 Komazawa, Setagaya-Ku, Tokyo 154-8525, Japan}
\email{takhshhr@komazawa-u.ac.jp}

\author[orcid=0000-0002-2309-3639, gname=Ken, sname='Ohsuga']{Ken Ohsuga} 
\affiliation{Center for Computational Science, University of Tsukuba, 1-1-1 Tennodai, Tsukuba, Ibaraki 305-8577, Japan}
\email{ohsuga@ccs.tsukuba.ac.jp}

\begin{abstract}
Active galactic nuclei (AGNs) commonly exhibit ultra-fast outflows (UFOs) with velocities of $0.03$--$0.3~c$. Recent high-resolution X-ray observations with XRISM/Resolve suggest that UFOs may have a clumpy structure. However, the formation and evolution of these clumpy outflows remain poorly understood. In the first paper of this series, we investigated the formation of clumpy outflows using numerical simulations and find that sufficiently high spatial resolution is required to reproduce clumpy structures numerically. In this second paper, we perform a series of numerical experiments to identify the key physical processes governing clump formation. We find that spatial fluctuations of physical quantities at the disk surface are essential for the formation of clumpy outflows, whereas spatial and temporal fluctuations at the inner boundary have only a minor effect. Furthermore, density structure-function analysis suggests that the high-density regions have an elongated morphology with characteristic transverse widths of several tens of $R_{\rm g}$, possibly produced by radiation Rayleigh--Taylor instability. A more comprehensive investigation of the instability in general-relativistic radiation magnetohydrodynamic simulations, where accretion disk structures emerge self-consistently, is expected to provide deeper insights into the physical origin of clumpy outflows.
\end{abstract}


\section{Introduction} 

Gas accretion onto compact objects is one of the most efficient mechanisms for converting gravitational potential energy into radiation. It powers a wide variety of energetic phenomena, including active galactic nuclei \citep[AGNs; e.g.,][]{Netzer+2013}, X-ray binaries \citep[XRBs; e.g.,][]{Done+2007}, tidal disruption events \citep[TDEs; e.g.,][]{Komossa+2015}, and gamma-ray bursts \citep[GRBs; e.g.,][]{Piran+2004}. 
Various black hole (BH) accretion models \citep[][]{Shakura+1973, Narayan+1994,Narayan+1995b,Abramowicz+1988,Narayan+2000, Quataert+2000, Narayan+2003, Tchekhovskoy+2011,Narayan+2012} have been proposed to account for different aspects of BH accretion at different rates. Although different models differ substantially in their disk structures and energy transport mechanisms, they all predict the presence of outflows driven by radiation pressure, magnetic fields, thermal expansion, or combinations of these processes \citep[e.g.,][]{Blandford+1982,Proga+2000,Ohsuga+2005,Takeuchi+2013,Yuan+2014,Hu+2022a,Yoshioka+2024}. Such outflows are thought to play a central role in regulating BH growth while transporting mass, momentum, and energy into the surrounding environment, thereby linking BH accretion to galaxy evolution through AGN feedback \citep{Fabian+2012,King+2015}. 

Observational evidence for accretion-driven outflows has accumulated across a broad range of systems, from stellar-mass BHs \citep{Kaaret+2017} to luminous quasars hosting supermassive BHs \citep[SMBHs;][]{King+2015}. Nevertheless, our understanding of these outflows remains limited by current observational capabilities. Key physical properties, including outflow velocity, ionization state, and mass-loss rate, are primarily inferred from the Doppler shifts, line widths, and profiles of absorption and emission features observed in X-ray and ultraviolet spectra \citep[e.g.,][]{Krolik+2001,Crenshaw+2003,Tombesi+2010,Gofford+2013,Laha+2021}. Meanwhile, numerical simulations have explored the launching and propagation of outflows under a variety of physical conditions \citep{Ohsuga+2005,Ohsuga+2009,Tchekhovskoy+2011,Sadowski+2014,Jiang+2014,Kitaki+2018,Hu+2022a,Yoshioka+2024,Hu+2026}, revealing a broad diversity of outflow properties spanning different spatial scales, ionization conditions, and acceleration mechanisms. However, directly connecting theoretical predictions with observational diagnostics remains a major challenge.

Recent advances in high-resolution X-ray spectroscopy have begun to reveal that accretion-driven outflows are considerably more structured than previously assumed. In particular, observations obtained with the \textit{X-Ray Imaging and Spectroscopy Mission} (XRISM) have provided unprecedented views of ultra-fast outflows in nearby AGNs \citep{Xrism+2025,Noda+2025,Mehdipour+2025,Gu+2025,Zaidouni+2026}. High-resolution spectroscopic observation of the luminous quasar PDS~456 \citep{Xrism+2025} indicates that its outflow exhibits pronounced spatial and temporal inhomogeneity, favoring a clumpy rather than homogeneous wind morphology. Similar evidence for clumpy or multiphase outflows has been reported in several AGNs and XRBs through variable absorption features, partial-covering absorbers, and rapid spectral variability \citep[e.g.,][]{Xrism+2025b,Ninoyu+2026}. These observations suggest that density fluctuations and localized structures are intrinsic characteristics of radiation-driven winds. Despite their importance, however, the physical origin, evolution, and statistical properties of clumpy outflows remain poorly understood. In particular, it is still unclear how radiative and hydrodynamic instabilities generate clumps \citep[see also][]{Takeuchi+2013}, how these clumps evolve as they propagate outwards, and how they influence both observational signatures and the efficiency of AGN feedback.

Understanding the formation and evolution of clumpy outflows is therefore essential for establishing a comprehensive picture of BH accretion and feedback. Motivated by these recent observational discoveries, we perform multidimensional radiation-hydrodynamic (RHD) simulations to investigate the physical properties of radiation-driven clumpy outflows. In the first paper of this series, we investigated the formation and statistical properties of clumpy outflows in numerical simulations \citep{Hu+2026}. 
We demonstrated that radiation-driven outflows from super-Eddington accretion disks naturally develop a clumpy structure, whose velocity, ionization states, and line-of-sight number of clumps are broadly consistent with recent XRISM observations. Building on these results, the present paper investigates the physical conditions required for clump formation and the hydrodynamic instabilities responsible for shaping the clumpy structures.

The remainder of this paper is organized as follows. Section~\ref{sec:method} describes the numerical methods, governing equations, and simulation setups. Section~\ref{sec:results} presents the results of our numerical experiments exploring different inner-boundary and disk-boundary conditions. Section~\ref{sec:discussion} discusses the dynamical structure functions and the roles of hydrodynamic instabilities in the formation of clumpy outflows. Finally, Section~\ref{sec:conclusion} summarizes our main conclusions.

\section{Method \& Equations} \label{sec:method}

In this paper, we perform two-dimensional radiation-hydrodynamic simulations to investigate the formation mechanism of clumpy outflows from super-Eddington accreting SMBHs using the UWABAMI+INAZUMA code \citep{Takahashi+2016,Asahina+2020,Asahina+2024}. The code incorporates a general-relativistic radiation magnetohydrodynamics (GR-RMHD) framework. However, general relativistic effects are not important in the present study. To isolate the hydrodynamic processes responsible for clump formation, magnetic fields are set to zero in all simulations. Previous studies have suggested that magnetic fields are not essential for the formation and evolution of clumpy outflows \citep{Takeuchi+2013}, although they may still contribute to the launching and acceleration of outflows. A detailed investigation of magnetic effects is beyond the scope of this paper and is deferred to future work.

\subsection{Key equations in UWABAMI+INAZUMA code}
 
 The numerical setup is identical to that presented in \citet{Hu+2026}. Here, we briefly summarize the key equations solved by the UWABAMI+INAZUMA code. Throughout the simulations, the speed of light is set to unity ($c=1$), although we restore $c$ where appropriate when discussing physical quantities. Following the standard convention in general relativity, Latin indices denote spatial components, whereas Greek indices denote spacetime components.
 
 The conservation equations of general relativistic radiation hydrodynamics are solved numerically in the simulations (see \citealt{Hu+2026} for a complete description). The gas energy-momentum tensor is given by
\begin{equation}
    T^\mu_\nu = \left ( \rho + \frac{\Gamma}{\Gamma -1}p\right )u^\mu u_\nu + p\delta^\mu_\nu,
\end{equation}
where $\rho$ is the gas density, $p$ is the gas pressure, $\delta^\mu_\nu$ is the Kronecker delta, and $\Gamma=5/3$ is the adiabatic index.
The interaction between gas and radiation is described by the radiation four-force $G_\nu$, which accounts for both absorption and electron scattering:
\begin{equation}
\begin{split}
    G_\nu = &-\rho \kappa_{\rm abs} (R_{\nu \alpha} u^{\alpha} + 4 \pi B u_\nu) \\
    &-\rho \kappa_{\rm sca} (R_{\nu \alpha} u^{\alpha} + R_{\alpha \beta} u^{\alpha}u^{\beta}u_{\nu}),
\end{split}
\end{equation}
where $\kappa_{\rm abs}=6.4\times10^{22}\rho T_{\rm gas}^{-7/2}\ {\rm cm^2~g^{-1}}$ and $\kappa_{\rm sca}=0.4\ {\rm cm^2~g^{-1}}$ are the absorption and electron-scattering opacities in the comoving frame \citep[see][]{Rybicki+1986}.
The blackbody intensity is given by
\begin{equation}
    B = \frac{a_{\rm rad} T_{\rm gas}^4}{4\pi},
\end{equation}
where $a_{\rm rad}$ is the radiation constant. The gas temperature $T_{\rm gas}$ is determined from the ideal-gas equation of state,
\begin{equation}
    p = \frac{\rho k_{\rm B}T_{\rm gas}}{\mu m_{\rm p}},
\end{equation}
where $k_{\rm B}$ is the Boltzmann constant, $m_{\rm p}$ is the proton mass, and $\mu=0.5$ is the mean molecular weight.
To close the radiation moment equations, we adopt the M-1 closure scheme \citep{Gonzalez+2007}. Under this approximation, the radiation energy-momentum tensor is written as
\begin{equation}
  R^{\mu\nu} = 4p_\mathrm{rad}u^{\mu}_\mathrm{rad}u^{\nu}_\mathrm{rad}+p_\mathrm{rad}g^{\mu\nu}, 
\end{equation}
where $p_{\rm rad}$ and $u_{\rm rad}^{\mu}$ denote the radiation pressure and the four-velocity of the radiation rest frame, respectively.

\subsection{Simulation setup \& Initial conditions}
\label{sec:IC}

In the simulations, we adopt spherical coordinates $(r,\theta,\phi)$ around a static, non-rotating BH with mass $M_{\rm BH}=10^7~M_\odot$. The computational domain extends from an inner radius of $30~R_{\rm g}$ to an outer radius of $1500~R_{\rm g}$, where $R_{\rm g}=GM_{\rm BH}/c^2$ is the gravitational radius. Although the simulations are performed in code units with $G=c=1$, physical units are restored throughout the discussion where appropriate. The computational domain covers polar angles of $5^\circ \leq \theta \leq 90^\circ$, excluding the vicinity of the polar axis to avoid contamination from the high-velocity jets. The domain is discretized using a uniform grid with $(N_r,N_\theta)=(640,315)$ in the radial and polar directions, respectively.

Following the procedure described in \citet{Hu+2026}, we adopt as the initial condition a snapshot at $t=4.9\times10^8~{\rm s}$ from the RHD simulations of \citet{Yoshioka+2024}. The RHD simulation data are converted into the GR-RMHD input format and bilinearly interpolated onto the computational grid where necessary. The resulting initial state consists of a turbulent, dense accretion disk, radiation-driven outflows above the disk surface (approximately between $\theta\sim10^\circ$ and the disk surface), and a high-velocity jet confined to the vicinity of the polar axis ($\theta\lesssim10^\circ$).

\subsection{Boundary conditions}
\label{sec:BC}

The computational domain corresponds to one quarter of a meridional annulus. Boundary conditions are prescribed in both the radial and polar directions. Along the polar direction, zero-gradient (outflow) boundary conditions are imposed on all physical quantities at $\theta=5^\circ$ and $\theta=90^\circ$. To isolate the dynamics of radiation-driven outflows from the evolution of the accretion disk, a disk boundary is additionally imposed throughout the simulations.

Following \citet{Hu+2026}, the disk boundary is constructed from the RHD simulations of \citet{Yoshioka+2024}. The disk surface is fitted by
\begin{equation}
    \theta=\theta_0 + \frac{1}{2}\cdot (\theta_1-\theta_0)\cdot \left (1+\tanh \left [  \frac{r-\delta r}{{\delta r}} \right] \right ),
\end{equation}
with $\theta_0=\pi/12$ and $\theta_1=1.4$ and $\delta r = 2000~R_{\rm g}$.
The fitted boundary closely follows the disk surface obtained by \citet{Yoshioka+2024}, where the radiation force approximately balances gravity. During the simulations, physical quantities along the disk boundary are obtained by bilinear interpolation between neighboring grid cells and linearly interpolated between two adjacent temporal snapshots. The boundary data are updated every $\delta t=10~R_{\rm g}/c$.

In the radial direction, outflow boundary conditions are imposed at the outer boundary ($R_{\rm out}=1500~R_{\rm g}$). At the inner boundary ($R_{\rm in}=30~R_{\rm g}$), all physical quantities are prescribed using the simulation data of \citet{Yoshioka+2024}, following the same procedure adopted for the disk boundary. Consequently, the effective computational domain is defined as the region enclosed by the inner and outer radial boundaries and located above the prescribed disk surface.

Among these boundary conditions, those imposed at the inner radial boundary ($R_{\rm in}=30~R_{\rm g}$) and along the disk surface are expected to have the greatest influence on the formation and evolution of clumpy outflows. In the following section, we investigate how spatially and temporally smoothed boundary conditions affect the development of clumpy structures.

\section{Numerical Experiments and Results} \label{sec:results}

The first emergence of clumpy outflows in our simulations was presented in \citet{Hu+2026}, where the numerical requirements for their formation were systematically examined. In this section, we perform a series of numerical experiments to investigate the physical mechanisms responsible for the formation of clumpy outflows. In Section~\ref{sec:IBC}, we examine the effects of the inner boundary conditions and their dependence on both spatial and temporal fluctuations. In Section~\ref{sec:DBC}, we investigate how different levels of smoothing applied to the disk boundary conditions influence the emergence of clumps. Finally, in Section~\ref{sec:HRDBC}, we conduct a broader parameter survey to assess the relative importance of different physical quantities. The simulation setups considered in this work are summarized in Table~\ref{tab:setup}.

\begin{deluxetable*}{llllllllr}
\digitalasset
\tablewidth{0pt}
\tablecaption{Summary of the numerical experiments performed in this work. \label{tab:setup}}
\tablehead{
\colhead{No.} & \colhead{ID Name} & \colhead{Inner BC} & \colhead{Disk BC-Rho} & \colhead{Vel.} & \colhead{Prs.} & \colhead{Rad.} & \colhead{Clumps} 
 & \colhead{Figure}
}
\startdata
1     & Fiducial                      		& Data              		& Data      	& Data      	& Data      	& Data      	& Strong  	& Fig.~\ref{fig:IBC-Check} \\
2     & IBC\_Smooth\_NoSteady       	& Smoothed          	& Data      	& Data      	& Data      	& Data      	& Strong  	& Fig.~\ref{fig:IBC-Check}\\
3     & IBC\_Smooth\_Steady           	& Smoothed+Std      & Data      		& Data      	& Data      	& Data      	& Strong  	& Fig.~\ref{fig:IBC-Check} \\
4     & DBC\_Seg\_5Rg                 	& Smoothed+Std      & 5Rg       		& 5Rg       	& 5Rg       	& 5Rg       	& Weak  	& Fig.~\ref{fig:Seg-Check}\\
5     & DBC\_Seg\_30Rg                	& Smoothed+Std      & 30Rg      	& 30Rg      	& 30Rg      	& 30Rg      	& Weak  	& Fig.~\ref{fig:Seg-Check}\\
6     & DBC\_Seg\_300Rg               	& Smoothed+Std      & 300Rg     	& 300Rg     	& 300Rg     	& 300Rg     	& None    	& Fig.~\ref{fig:Seg-Check}\\
7     & DBC\_Rho\_Smooth              	& Smoothed+Std      & Smoothed  	& Data      	& Data      	& Data      	& Weak   	& Fig.~\ref{fig:DBCRho-Check}\\
8     & DBC\_Rho\_No\_Smooth     	& Smoothed+Std      & Data      		& Smoothed  	& Smoothed  	& Smoothed  	& Weak   	& Fig.~\ref{fig:DBCRho-Check}\\
9     & DBC\_Rho\_Smooth\_IBC    	& Data              		& Smoothed  	& Data      	& Data      	& Data      	& Moderate  	& Fig.~\ref{fig:DBCRho-Check}\\
10   & DBC\_Vel\_Smooth              	& Smoothed+Std      & Data      		& Smoothed  	& Data      	& Data      	& Weak   	& Fig.~\ref{fig:RadHydro1}, \ref{fig:RadHydro2}\\
11   & DBC\_Vel\_No\_Smooth          & Smoothed+Std      & Smoothed  	& Data      	& Smoothed  	& Smoothed  	& Weak  	& Fig.~\ref{fig:RadHydro1}, \ref{fig:RadHydro2}\\
12   & DBC\_Prs\_Smooth              	& Smoothed+Std      & Data      		& Data      	& Smoothed  	& Data      	& Weak 	& Fig.~\ref{fig:RadHydro1}, \ref{fig:RadHydro2}\\
13   & DBC\_Prs\_No\_Smooth        	& Smoothed+Std      & Smoothed  	& Smoothed  	& Data      	& Smoothed  	& Weak 	& Fig.~\ref{fig:RadHydro1}, \ref{fig:RadHydro2}\\
14   & DBC\_Rad\_Smooth              	& Smoothed+Std      & Data      		& Data      	& Data      	& Smoothed  	& Weak 	& Fig.~\ref{fig:RadHydro1}, \ref{fig:RadHydro2}\\
15   & DBC\_Rad\_No\_Smooth      	& Smoothed+Std      & Smoothed  	& Smoothed  	& Smoothed  	& Data      	& Weak 	& Fig.~\ref{fig:RadHydro1}, \ref{fig:RadHydro2}\\
\enddata
\tablecomments{
Summary of the numerical experiments performed in this work. Columns (3)--(7) specify the treatment of the inner boundary condition (Inner BC) and the disk boundary conditions for density (Rho), velocity (Vel.), gas pressure (Prs.), and radiation field (Rad.), respectively. ``Data'' indicates that the boundary conditions are directly adopted from the radiation-hydrodynamic simulations of \citet{Yoshioka+2024}, whereas ``Smoothed'' denotes spatially (segment-averaged over $30\,R_{\rm g}$) and temporally averaged boundary conditions. For the segmented boundary-condition models, ``5Rg'', ``30Rg'', and ``300Rg'' represent radial averaging lengths of $5\,R_{\rm g}$, $30\,R_{\rm g}$, and $300\,R_{\rm g}$, respectively. The final two columns qualitatively summarize the degree of clump formation in each simulation and corresponding Figures.
}
\end{deluxetable*}

\subsection{Inner Boundary Conditions: spatial \& temporal fluctuations}\label{sec:IBC}

In \citet{Hu+2026}, we demonstrated that resolving the inner accretion flow ($r\lesssim300~R_{\rm g}$) is essential for reproducing clumpy outflows, based on simulations with different inner boundary locations ($R_{\rm in}=30$-$300~R_{\rm g}$). Motivated by this result, we fix the inner boundary at $R_{\rm in}=30~R_{\rm g}$ throughout this work and investigate how the spatial and temporal variability of the inner boundary conditions influences the formation of clumpy outflows.

As a control simulation, the ``Fiducial'' model (Row 1 of Table~\ref{tab:setup}) adopts the same setup as the fiducial model in \citet{Hu+2026}. For consistency, clumpy outflows are selected using the same criterion based on the ionization parameter,
\begin{equation}
\xi = \frac{4\pi F_{r}}{\rho/\mu m_{\rm p}}
\label{eq:ionizationparametert}
\end{equation}
where we require $3.0\lesssim\log_{10}(\xi/[{\rm erg~s^{-1}~cm}])\lesssim5.0$ \citep[see][for alternative selection criteria]{Hu+2026}.

To examine the role of the inner boundary, we perform two additional simulations (Rows 2 and 3 of Table~\ref{tab:setup}) using ``Smoothed'' and ``Steady'' inner boundary conditions. For the ``Smoothed'' model, all physical quantities are averaged over the polar-angle range $5^\circ\le\theta\le45^\circ$, where $\theta=45^\circ$ approximately corresponds to the disk surface at the inner boundary. This procedure removes spatial fluctuations along the polar direction. We note that such averaging reduces the radial velocity near the polar axis while occasionally enhancing it close to the disk surface. As a result, the smoothing process slightly smears out the time-fluctuations of outflow rates at the inner boundary compared to actual RHD data.
For the ``Steady'' model, a single snapshot is used to prescribe the boundary condition, thereby eliminating temporal variability.
The configurations of these three simulations are summarized in Rows 1-3 of Table~\ref{tab:setup}. 

In the Fiducial model, the boundary conditions are directly adopted from the RHD simulations (``Data'' in Table~\ref{tab:setup}). The ``IBC\_Smooth\_NoSteady'' model employs spatially smoothed boundary conditions while retaining their temporal variability, whereas the ``IBC\_Smooth\_Steady'' model further suppresses temporal variability by adopting a single snapshot (``Steady'', abbreviated as ``Std''). All simulations reach a quasi-steady state after several dynamical timescales ($t_{\rm dyn}\sim R_{\rm out}/c$). Unless otherwise stated, we present snapshots at $t=1.6\times10^5~t_{\rm g}$, while time-averaged quantities are computed over $t=(0.9$-$1.7)\times10^5~t_{\rm g}$. The data-analysis procedure follows that of \citet{Hu+2026}.

\begin{figure*}[tb]
\centering
\includegraphics[width=0.95\linewidth]{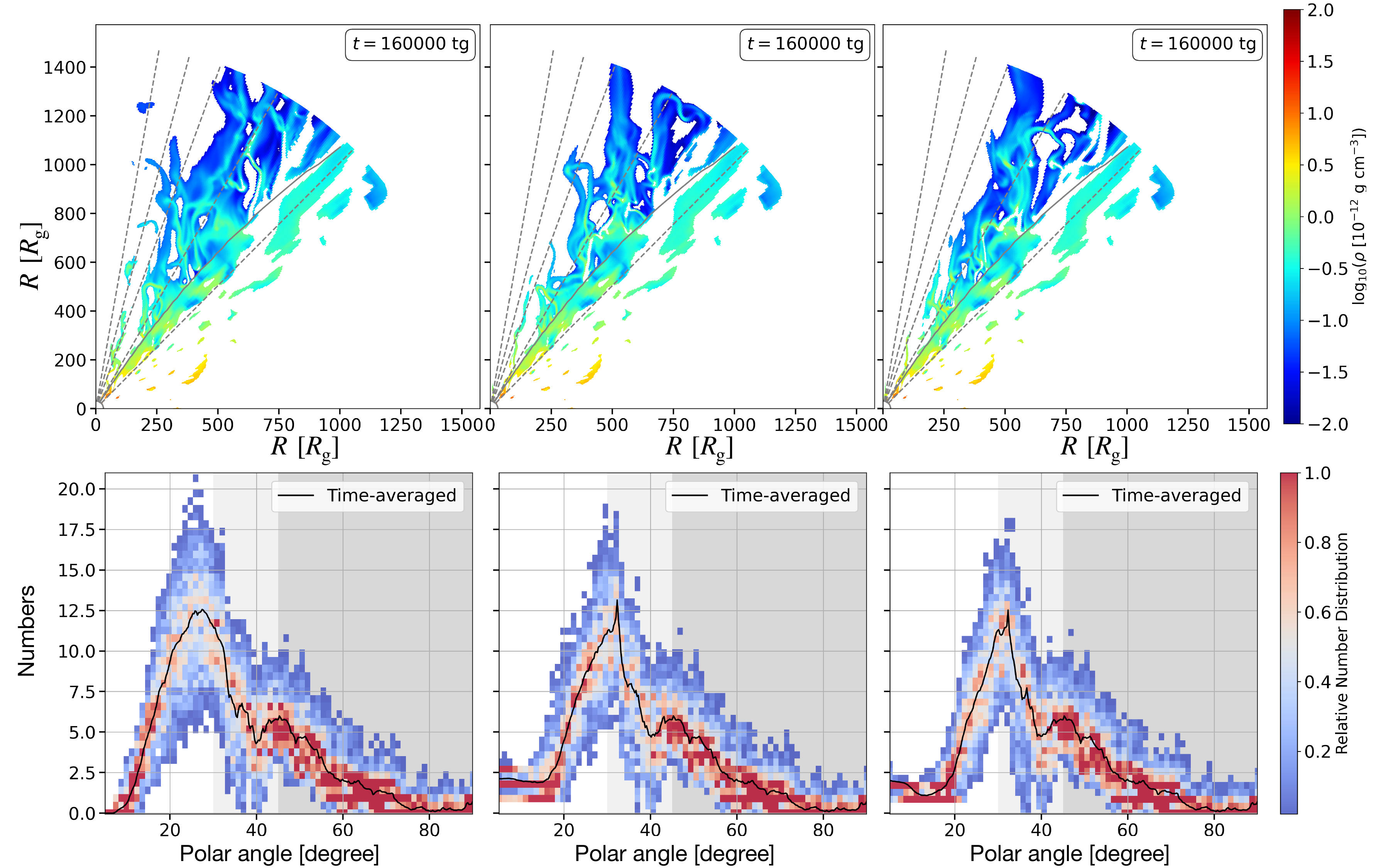}
\caption{
Upper panels: Two-dimensional density distributions of the selected clumpy outflows for the Fiducial, IBC\_Smooth\_NoSteady, and IBC\_Smooth\_Steady simulations (left to right). Clumps are selected according to the ionization criterion described in \citet{Hu+2026}. Dashed gray curves indicate five representative lines of sight ($9^\circ$, $14^\circ$, $19^\circ$, $29^\circ$, and $44^\circ$), while the solid gray curve marks the disk boundary. Lower panels: Angular distributions of the number of clumps along each line of sight. The colored regions represent the distributions from individual snapshots spanning $t=(0.9–1.7)\times10^5~t_{\rm g}$ with a cadence of $\delta t=200~t_{\rm g}$, whereas the black curve shows the time-averaged distribution. For $30^\circ<\theta<45^\circ$, the lines of sight partially intersect the disk (light-gray region), while for $\theta>45^\circ$ they are completely embedded within the disk (gray region).
\label{fig:IBC-Check}}
\end{figure*}

Figure~\ref{fig:IBC-Check} presents the morphology and statistical properties of the clumpy outflows. The upper panels show the two-dimensional density distributions of the selected clumps, while the lower panels display the angular distributions of the clump number along different line-of-sights. In the Fiducial model (first column), high-velocity outflows launched from the inner boundary propagate outward and interact with the slower disk winds, leading to the formation of clumpy structures \citep[see also][]{Hu+2026}. Except for the immediate vicinity of the polar axis, clumps develop over a broad range of polar angles. On average, more than five clumps are found between $\theta\sim10^\circ$ and $\theta\sim40^\circ$, corresponding to a covering factor of approximately 0.44 \citep{Hu+2026}.

The second column shows the results of the IBC\_Smooth\_NoSteady model. Although the overall morphology remains similar to that of the Fiducial model, noticeable differences are evident. The angular distribution exhibits a lower and narrower peak than that of the Fiducial model, indicating that fewer clumps form at larger polar angles. This behavior is attributed to the spatial smoothing of the inner boundary conditions. Near the polar axis, the flow is dominated by high-velocity gas originating from the inner boundary. Smoothing the boundary conditions reduces angular inhomogeneities in this fast outflow, allowing it to erase small-scale structures more efficiently as it propagates outward. This effect is most pronounced near the polar region, where the interaction with the dense, slower disk winds is relatively weak. These results suggest that spatially smoothed inner boundary conditions do not prevent clump formation, though they modestly suppress the development of clumpy structures in the polar outflow.

In the IBC\_Smooth\_Steady model (third column), we further eliminate the temporal variability of the inner boundary conditions by prescribing a single boundary snapshot throughout the simulation. Consequently, neither spatial nor temporal fluctuations are present at the inner boundary. Despite this additional simplification, the statistical properties remain remarkably similar to those of the IBC\_Smooth\_NoSteady model, both exhibiting a lower and narrower angular distribution than the Fiducial model. 
This comparison suggests that neither the spatial nor temporal fluctuations in the inner boundary conditions are essential for clump formation, although they do have a modest influence on the detailed properties of the clumpy outflows. 
It is worth noting that as long as the spatial fluctuations at the disk boundary are preserved, the fluctuations at the inner boundary have little effect on clump formation (see also Section \ref{sec:density}).

\subsection{Segment-averaged Disk Boundary Conditions}\label{sec:DBC}

We next investigate the influence of spatial fluctuations in the disk boundary conditions. To this end, we perform a series of simulations in which the disk boundary conditions are smoothed over different spatial scales (hereafter ``segment-averaged''), while the inner boundary conditions are kept spatially smoothed and temporally steady.

In these experiments, we compute arithmetic averages of individual physical quantities (e.g., density, velocity, gas pressure, and each component of the radiation energy–momentum tensor) over prescribed radial segments along the disk surface.\footnote{Averaging over the entire radial extent would introduce substantial errors because many physical quantities exhibit strong global radial gradients. The segment-averaged method preserves the large-scale radial dependence while smoothing small-scale spatial fluctuations.} Three segment lengths, $S_{\rm seg}=5~R_{\rm g}$, $30~R_{\rm g}$, and $300~R_{\rm g}$, are considered, corresponding to approximately 2, 10, and 100 times the radial grid spacing, respectively. This procedure suppresses spatial fluctuations on the chosen segment scale while retaining the large-scale radial dependence naturally expected in accretion-disk solutions such as the standard and slim disks.

\begin{figure}[tb]
\centering
\includegraphics[width=0.95\linewidth]{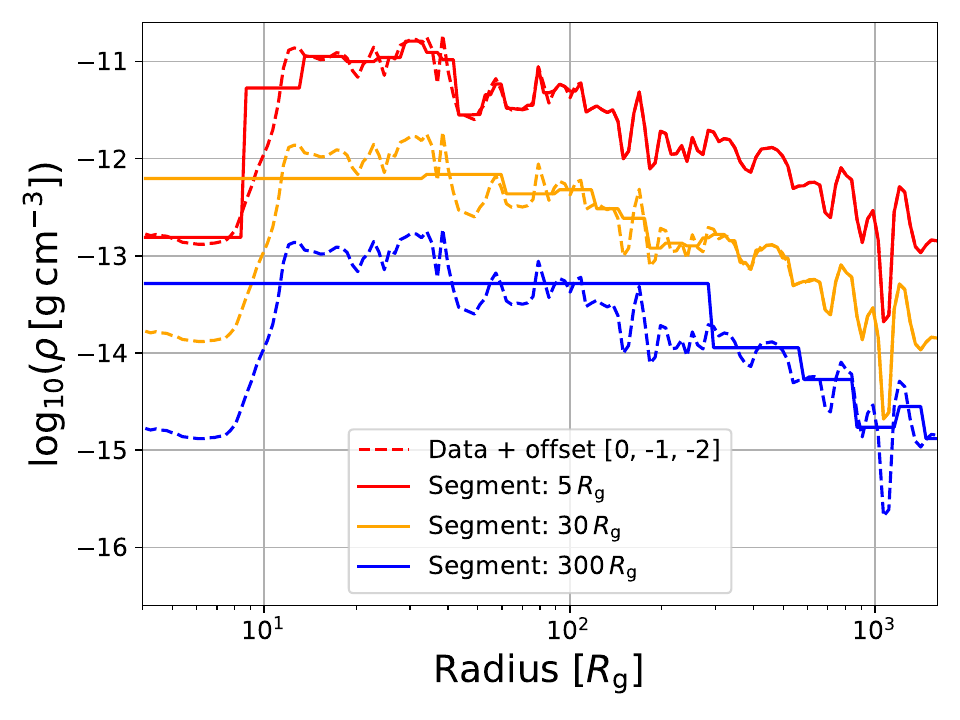}
\caption{
Comparison of the radial density profiles between the original radiation-hydrodynamic data (dashed curves) and the segment-averaged profiles (solid curves) for $S_{\rm seg}=5~R_{\rm g}$, $30~R_{\rm g}$, and $300~R_{\rm g}$. Vertical offsets of $-1$ and $-2$ are applied to the orange and blue curves for clarity. Increasing the segment length progressively suppresses small-scale spatial fluctuations while preserving the large-scale radial dependence.
\label{fig:SegmentAverage}}
\end{figure}

As an illustration, Figure~\ref{fig:SegmentAverage} compares the radial density profiles of the original RHD data with the corresponding segment-averaged profiles. The raw simulation data are shown as dashed curves, while the segment-averaged profiles are plotted as solid curves for $S_{\rm seg}=5~R_{\rm g}$, $30~R_{\rm g}$, and $300~R_{\rm g}$. Different vertical offsets are applied for clarity. As expected, increasing the segment length progressively suppresses small-scale spatial fluctuations while preserving the overall radial trend. Even for the largest segment ($S_{\rm seg}=300~R_{\rm g}$), the global radial dependence remains well reproduced.

\begin{figure*}[htb]
\centering
\includegraphics[width=0.95\linewidth]{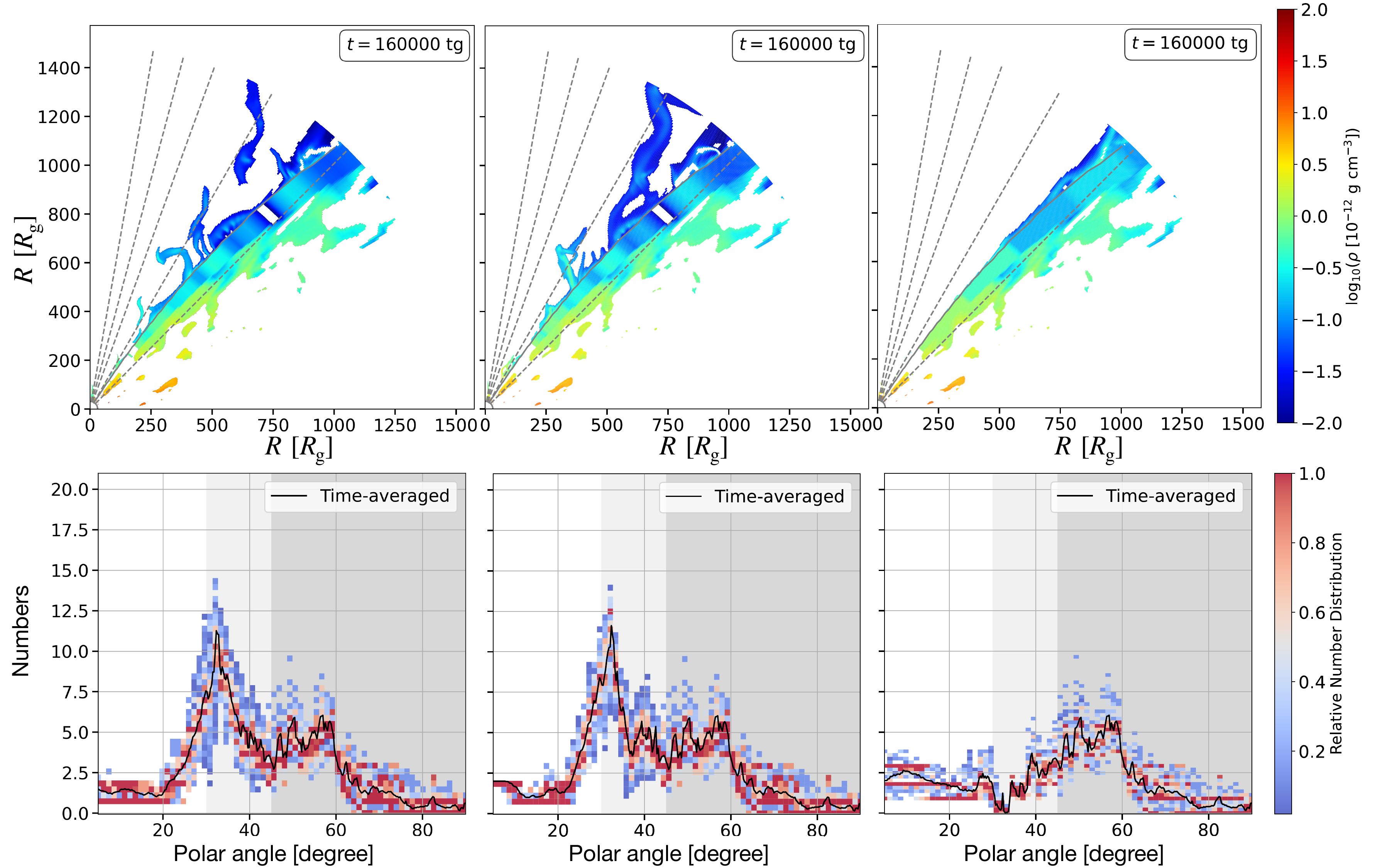}
\caption{
Same as Figure~\ref{fig:IBC-Check}, but for the DBC\_Seg\_5Rg, DBC\_Seg\_30Rg, and DBC\_Seg\_300Rg simulations (left to right). 
To illustrate the imposed disk boundary conditions, a thick layer beneath the disk surface is overlaid with the corresponding boundary values.
\label{fig:Seg-Check}}
\end{figure*}

Figure~\ref{fig:Seg-Check} presents the morphology and statistical properties of the clumpy outflows for the three segment-averaged models: DBC\_Seg\_5Rg, DBC\_Seg\_30Rg, and DBC\_Seg\_300Rg (Rows 4-6 of Table~\ref{tab:setup}).
The upper panels display the selected clumps at $t=1.6\times10^5~t_{\rm g}$. A clear trend emerges as the segment length increases: clumpy structures become progressively weaker and eventually disappear. In the left and middle panels, a number of clumps remain at $r\gtrsim200~R_{\rm g}$ along lines of sight around $\theta\sim30^\circ$. In contrast, almost no clumps are visible in the right panel, where the disk boundary conditions have been smoothed over the largest spatial scale. This trend is even more evident in the lower panels, which show the angular distributions of the clump number. For the DBC\_Seg\_5Rg and DBC\_Seg\_30Rg models, a pronounced peak of approximately 10 clumps remains near $\theta\sim30^\circ$, whereas no comparable peak is present in the DBC\_Seg\_300Rg model.

Despite the relatively mild smoothing applied in the DBC\_Seg\_5Rg model, noticeable differences from the Fiducial model (Figure~\ref{fig:IBC-Check}) are already apparent in both the morphology of the clumpy outflows and the angular clump-number distribution. Although the smoothing scale is only about twice the radial grid spacing, fewer clumps are produced and they become more concentrated toward the disk surface. These numerical experiments therefore indicate that spatial fluctuations along the disk boundary are essential for the formation of clumpy outflows. Even modest smoothing substantially alters the outflow morphology and statistical properties. Interestingly, the resulting suppression of clump formation resembles the resolution dependence reported in \citet{Hu+2026}, although the underlying physical mechanisms are fundamentally different, as discussed in Section 4.

\subsection{Hydro \& Radiation Boundary Conditions}\label{sec:HRDBC}

The previous subsection demonstrates that spatial fluctuations along the disk boundary play a crucial role in the formation of clumpy outflows. We now investigate which physical quantities are primarily responsible for this behavior by separately examining the effects of hydrodynamic variables (density, velocity, and gas pressure) and radiation quantities (the radiation stress-energy tensor). In Section~\ref{sec:density}, we focus on the role of density fluctuations, while Section~\ref{sec:others} examines the effects of the remaining physical quantities, including velocity, gas pressure, and the radiation stress-energy tensor. Unless otherwise specified, all smoothed quantities are segment-averaged over a radial length of $S_{\rm seg}=30~R_{\rm g}$.

\subsubsection{fluctuations in densities}\label{sec:density}

\begin{figure*}[tb]
\centering
\includegraphics[width=0.95\linewidth]{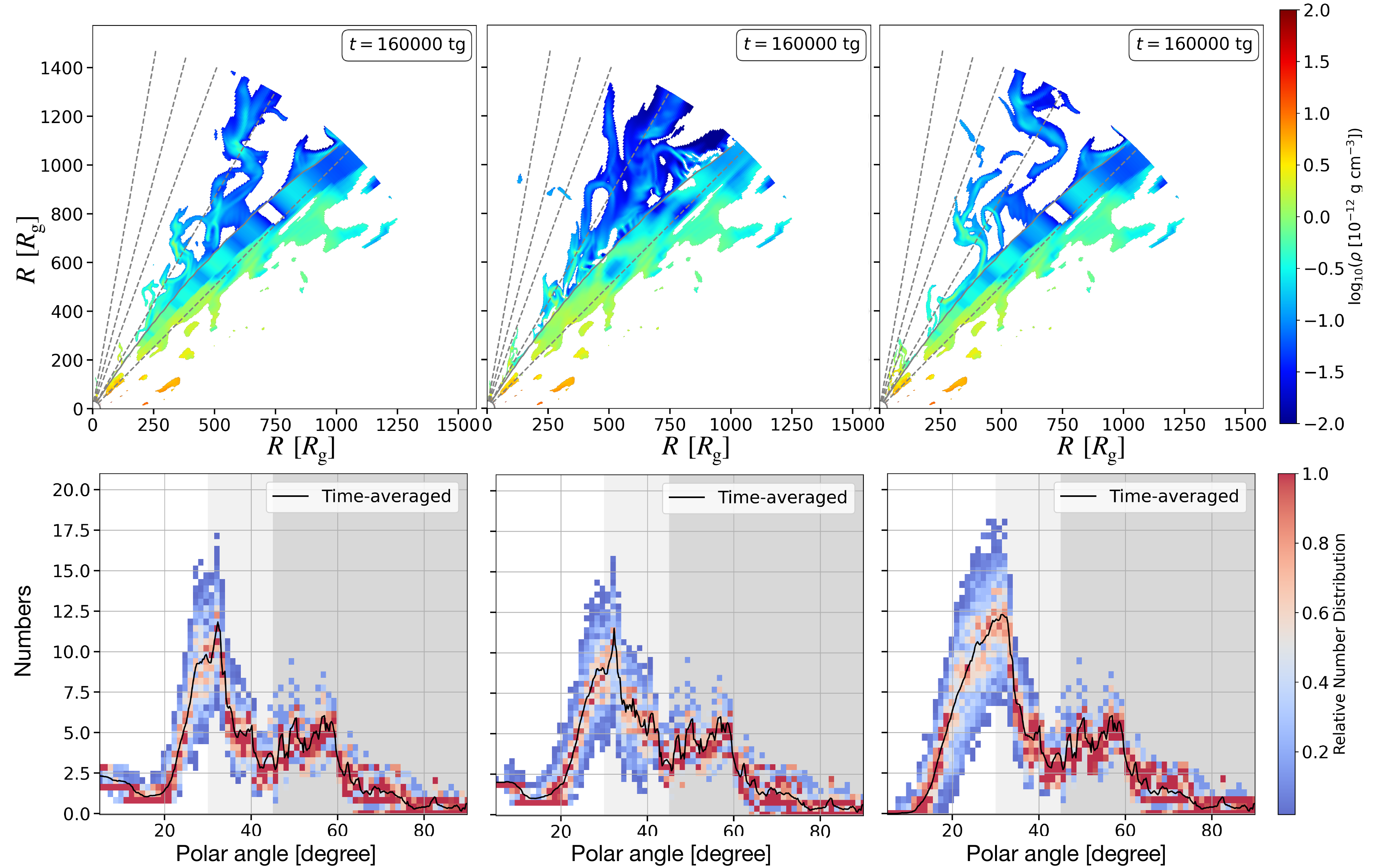}
\caption{
Same as Figure~\ref{fig:IBC-Check}, but for the DBC\_Rho\_Smooth, DBC\_Rho\_No\_Smooth, and DBC\_Rho\_Smooth\_IBC simulations (left to right).
\label{fig:DBCRho-Check}}
\end{figure*}

To investigate the role of density fluctuations, we perform three numerical experiments summarized in Rows 7-9 of Table~\ref{tab:setup}. Figure~\ref{fig:DBCRho-Check} presents the corresponding clump morphologies and angular clump-number distributions. The left panels show the DBC\_Rho\_Smooth model, in which only the density at the disk surface is smoothed over a segment length of $S_{\rm seg}=30~R_{\rm g}$, while all other quantities are directly adopted from the RHD simulations. The middle panels correspond to the DBC\_Rho\_No\_Smooth model, in which only the density retains its original spatial fluctuations. Comparing these two models, we find no significant differences in the morphology of the clumpy outflows, and their angular clump-number distributions remain remarkably similar. However, compared with the DBC\_Seg\_30Rg model (middle panels of Figure~\ref{fig:Seg-Check}), both models exhibit broader
peaks in the angular distribution, while maintaining a similar height. These results indicate that preserving the full spatial fluctuations of density along disk surface
promotes clump formation over a wider range of polar angles.

As a complementary test, we perform an additional simulation based on the DBC\_Rho\_Smooth model, but restore the full spatial fluctuations of all physical quantities at the inner boundary. This model is referred to as DBC\_Rho\_Smooth\_IBC. Compared with the DBC\_Rho\_Smooth model, the angular clump-number distribution exhibits a broader and higher peak, indicating enhanced clump formation, particularly toward smaller polar angles. This behavior is consistent with the results presented in Section~\ref{sec:IBC}, where the full spatial variability of the inner boundary conditions was shown to promote clump formation in the polar outflow. Overall, these numerical experiments suggest that preserving the spatial fluctuations of at least one physical quantity at the disk surface is sufficient to facilitate the formation of clumpy outflows, whereas restoring the full spatial fluctuations at the inner boundary almost has no effects in the formation of clumps.

\subsubsection{fluctuations in other quantities}\label{sec:others}

We next extend our analysis to the remaining physical quantities at the disk boundary, namely the radial velocity, gas pressure, and radiation stress-energy tensor. For each quantity, we perform two simulations in which only the selected quantity is either spatially smoothed or left unchanged. The corresponding simulation setups are summarized in Rows 10-15 of Table~\ref{tab:setup}.

\begin{figure*}[tb]
\centering
\includegraphics[width=0.96\linewidth]{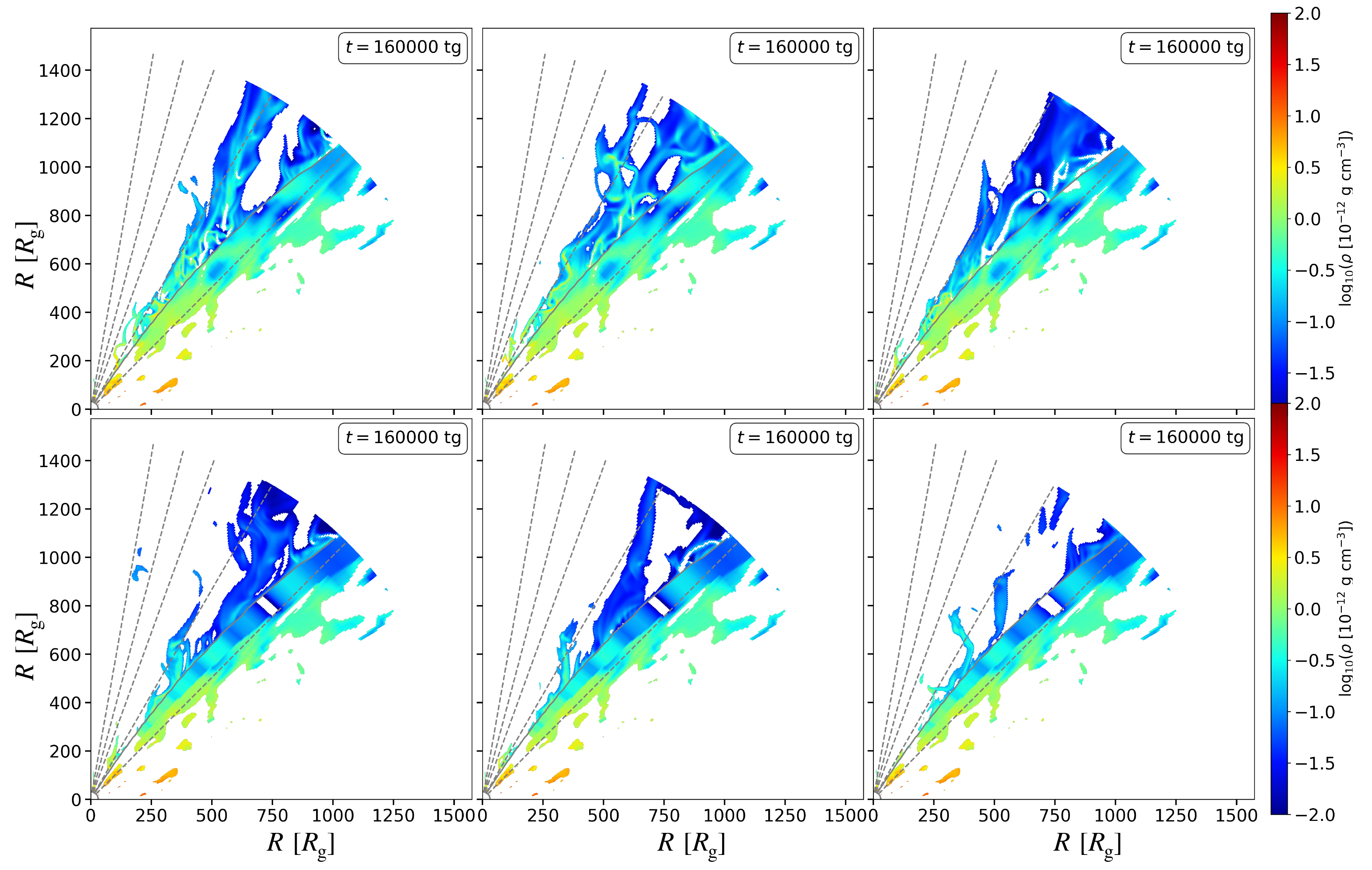}
\caption{
Same as the upper panels of Figure~\ref{fig:IBC-Check}, but for the DBC\_Vel\_Smooth, DBC\_Prs\_Smooth, and DBC\_Rad\_Smooth models (upper row, left to right), and the DBC\_Vel\_No\_Smooth, DBC\_Prs\_No\_Smooth, and DBC\_Rad\_No\_Smooth models (lower row, left to right).
\label{fig:RadHydro1}}
\end{figure*}

Figure~\ref{fig:RadHydro1} compares the two-dimensional distributions of clumpy outflows for the six simulations. The upper panels correspond to models in which only one physical quantity is spatially smoothed (from left to right: radial velocity, gas pressure, and radiation stress-energy tensor), whereas the lower panels show the complementary models in which only the corresponding quantity retains its original spatial fluctuations. Clumpy outflows are reproduced in all six simulations.
Compared with the upper panels, the lower panels exhibit a modest reduction in the number of clumps, consistent with a weaker degree of fragmentation.

\begin{figure*}[tb]
\centering
\includegraphics[width=0.95\linewidth]{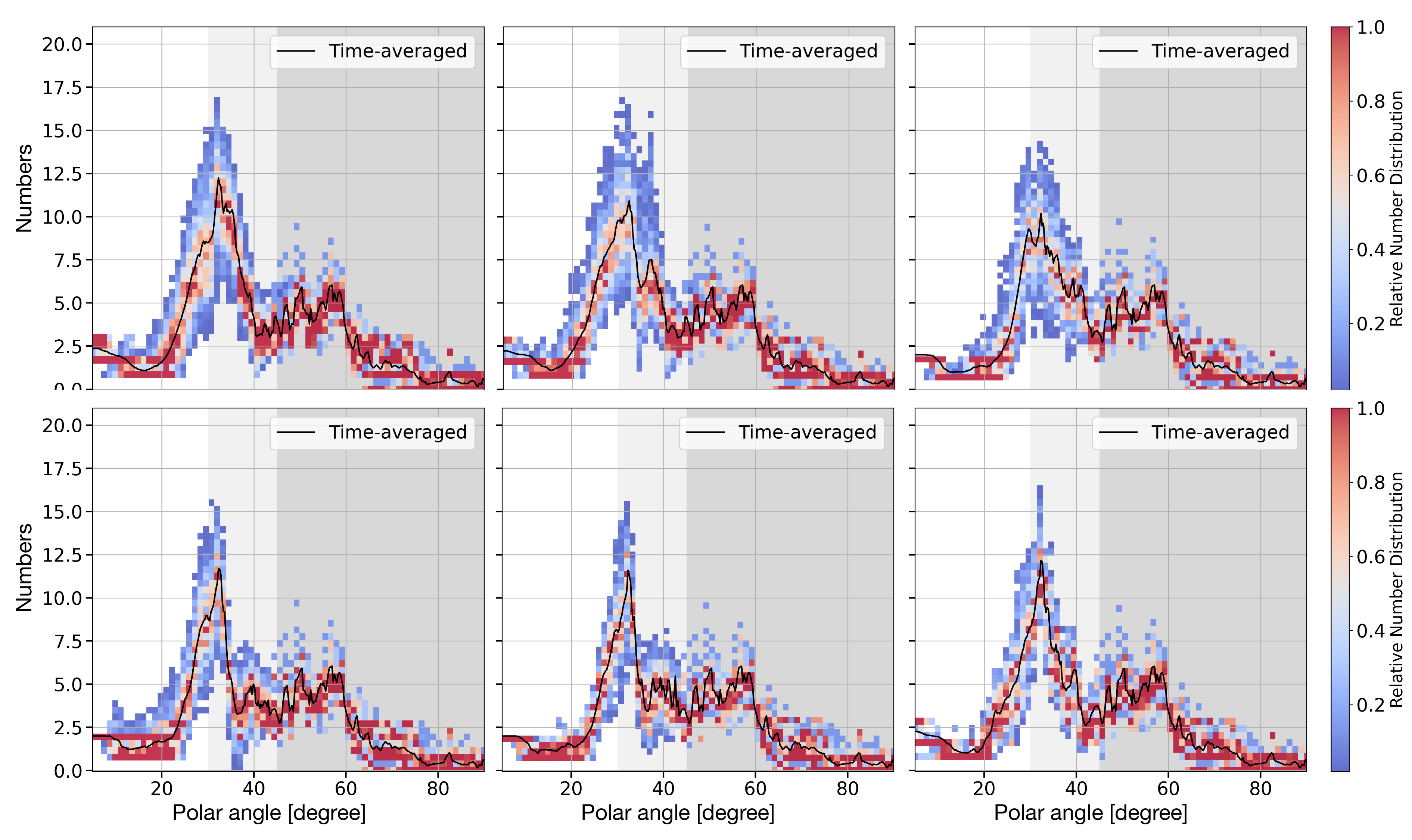}
\caption{
Same as the lower panels of Figure~\ref{fig:IBC-Check}, but for the DBC\_Vel\_Smooth, DBC\_Prs\_Smooth, and DBC\_Rad\_Smooth models (upper row, left to right), and the DBC\_Vel\_No\_Smooth, DBC\_Prs\_No\_Smooth, and DBC\_Rad\_No\_Smooth models (lower row, left to right).
\label{fig:RadHydro2}}
\end{figure*}

For a quantitative comparison, Figure~\ref{fig:RadHydro2} presents the angular clump-number distributions for all six simulations. In the upper panels, where only one physical quantity is spatially smoothed, the distributions exhibit similar peak amplitudes and widths, closely resembling those of the DBC\_Seg\_30Rg model (middle panel of Figure~\ref{fig:Seg-Check}) and the first two density models (left and middle panels of Figure~\ref{fig:DBCRho-Check}). This consistency further supports the conclusion that preserving the spatial fluctuations of at least one physical quantity at the disk boundary is sufficient to sustain clump formation. In contrast, the distributions in the lower panels are systematically narrower and lower, indicating fewer clumps and a reduced angular extent. This trend is consistent with the morphologies shown in Figure~\ref{fig:RadHydro1}. Taken together, these experiments indicate that increasing the number of physical quantities retaining their spatial fluctuations—either at the disk boundary or the inner boundary—promotes the formation of more numerous and more broadly distributed clumps.

\section{Discussion} \label{sec:discussion}
\subsection{Signatures for Instabilities}

In this section, we seek signatures of the physical mechanisms responsible for the formation of clumpy outflows. Radiation is clearly indispensable for clump formation, as simulations without radiative transfer produce only smooth outflows driven by gas pressure.\footnote{The simulation without the radiation module is presented in Appendix~\ref{app:A}.} Under super-Eddington accretion, thermal instability, radiation Rayleigh-Taylor (RT) instability driven by outward radiative acceleration, and radiation Kelvin-Helmholtz (KH) instability are potential candidates for driving clump formation. We first examine the possibility of thermal instability and then use structure-function analysis to investigate the spatial structure of the clumpy outflow and assess the possible roles of RT and KH instabilities.

In the simulation, gas above the disk is heated to temperatures of $T_{\rm gas}\gtrsim10^{6-7}~{\rm K}$, while radiation temperature is sustained around $T_{\rm rad}\sim10^5~{\rm K}$. This temperature difference naturally lead to net cooling due to radiative processes
\begin{equation}
Q_{\rm rad}^{-/+}= \rho_{\rm gas}\kappa_{\rm abs} a_{\rm rad} c
\left( T_{\rm gas}^4-T_{\rm rad}^4 \right),
\end{equation}
where $a_{\rm rad}$ is the radiative constant.
The radiative cooling timescale is
\begin{equation}
t_{\rm cool} = \frac{e_{\rm gas}}{Q^-_{\rm rad}}=\frac{\rho_{\rm gas} k_{\rm B} T_{\rm gas}}{\mu m_{\rm p}(\Gamma-1)\cdot Q^-_{\rm rad}}.
\end{equation}
For characteristic gas properties, the cooling timescale is
\begin{equation}
t_{\rm cool}\sim 2\times 10^6 ~{\rm s} \cdot \left(\frac{T_{\rm gas}}{10^8~{\rm K}}\right )^{1/2} \left (\frac{\rho_{\rm gas}}{10^{-13}~{\rm g~cm^{-3}}}\right)^{-1},
\end{equation}
which is somewhat longer than the dynamical/orbital timescale $t_{\rm dyn}=R/v_{\rm out}\sim 3\times10^5~{\rm s}$ at $R\sim300~R_{\rm g}$ for outflows with $v_{\rm out}\sim0.05~c$. This suggests that thermal instability does not develop significantly within the simulation domain. We note that the radiation temperature $T_{\rm rad}$ used in estimating the radiative cooling rate assumes blackbody radiation, whereas the actual photon energies may be higher than implied by this temperature. In that case, the cooling timescale would be even longer.

We next consider radiation RT and KH instabilities as possible mechanisms for producing density inhomogeneities in the outflow. In hydrodynamics, RT instability develops when a dense fluid is accelerated into a lighter fluid. This configuration naturally arises in our simulations, where the dense accretion disk underlies the diffuse outflowing gas and the outward radiation force provides the effective acceleration. In addition, velocity shear between neighboring regions of the outflow may allow radiation KH instability to grow. Both instabilities can therefore potentially generate or amplify density inhomogeneities.

\begin{figure*}[tb]
\centering
\includegraphics[width=0.98\linewidth]{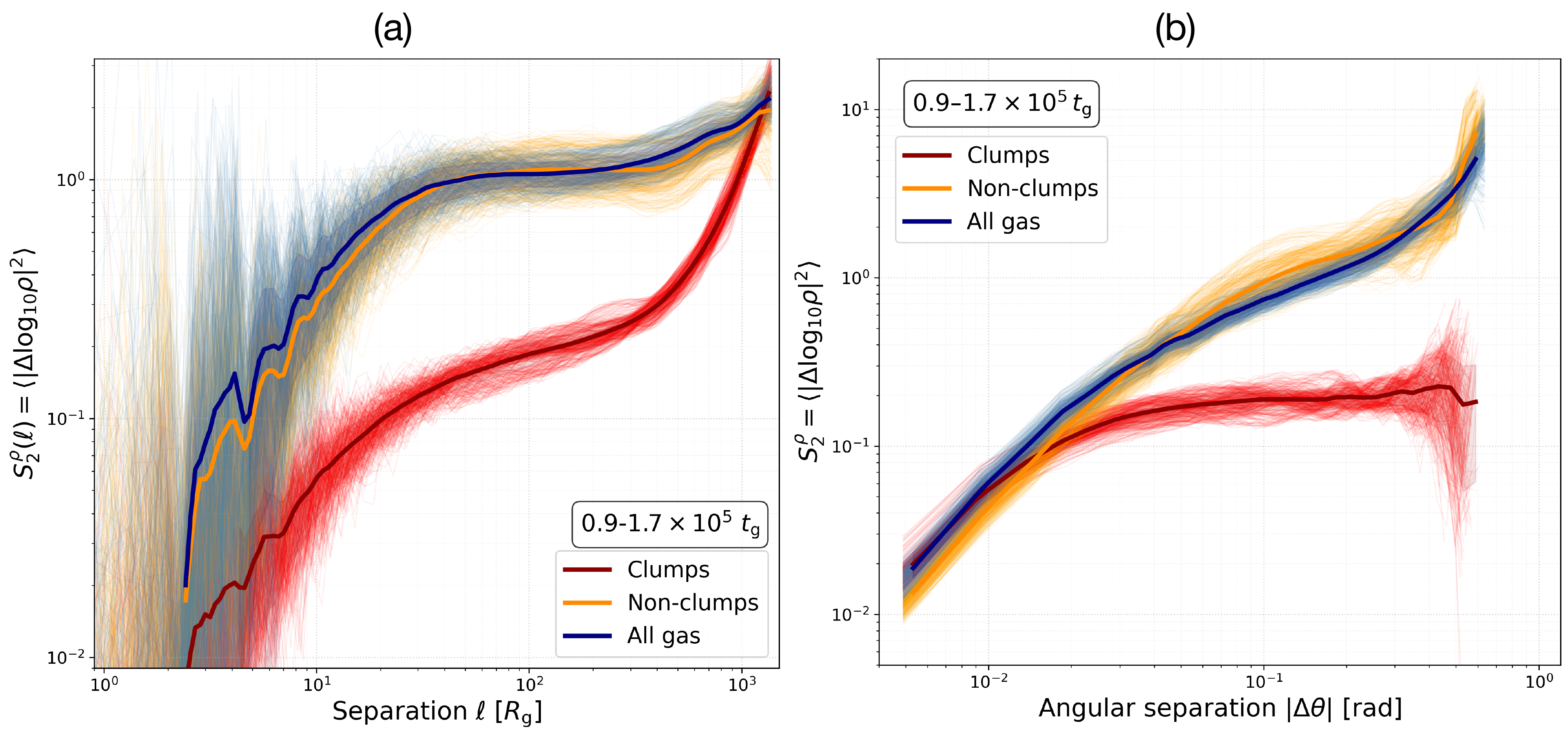}
\caption{
Left (a): the unrestricted second-order density structure functions for clumps (red), non-clumps (orange), and the entire gas distribution (blue). Thin curves show individual snapshots over the interval $t=(0.9$-$1.7)\times10^5~t_{\rm g}$ with a cadence of $\Delta t=200~t_{\rm g}$, while the thick curves denote the corresponding time-averaged structure functions. 
Right (b): the restricted second-order density structure functions measured at each fixed radius for clumps (red), non-clumps (orange), and the entire gas distribution (blue). Thin curves show individual snapshots over the interval $t=(0.9$-$1.7)\times10^5~t_{\rm g}$ with a cadence of $\Delta t=200~t_{\rm g}$, while the thick curves denote the corresponding time-averaged structure functions. 
\label{fig:StructureFunctions}}
\end{figure*}

To characterize the spatial structure of the outflow, we analyze the second-order density structure function. The second-order structure function is defined as
\begin{equation}
S_{2}^{m}(l) = \langle \left[ m(\vec{x}_1)-m(\vec{x}_2) \right]^2\rangle.
\end{equation}
where $l\equiv|\vec{x}_1-\vec{x}_2|$ is the separation between two cells at positions $\vec{x}_1$ and $\vec{x}_2$, and $m(\vec{x}_1)$ and $m(\vec{x}_2)$ are physical quantities at the two points, respectively. This definition measures the averaged difference at fixed separation $l$ and thereby characterizes spatial structures in the system. Figure~\ref{fig:StructureFunctions} shows the density structure function, $m=\log_{10}(\rho/10^{-12}~{\rm g~cm^{-3}})$, as a function of $l$ in panel (a), and the angular density structure function as a function of $\Delta\theta$ in panel (b). Red, orange, and blue curves denote clumps, non-clumps, and all gas components, respectively. Thin curves show individual snapshots, while thick curves show the corresponding time-averaged structure functions.

We first focus on the density structure function shown in Figure~\ref{fig:StructureFunctions}(a). For the clump regions, the structure function increases rapidly over $2\lesssim l/R_{\rm g}\lesssim30$, shows a broad plateau over $30\lesssim l/R_{\rm g}\lesssim500$, and increases again at larger separations. We therefore identify two characteristic scales, $l_1\sim30~R_{\rm g}$ and $l_2\sim500~R_{\rm g}$. The density structure functions of the non-clump regions and the entire outflow show broadly similar scale-dependent behavior, although their amplitudes differ from that of the clump regions. Since the clump regions are selected according to density, part of the difference in the density-structure-function amplitude may arise from this selection.

We next examine the angular density structure function shown in Figure~\ref{fig:StructureFunctions}(b) to clarify the geometrical meaning of the smaller characteristic scale $l_1$. For the clump regions, the angular structure function increases up to $\Delta\theta\sim0.02$--$0.04$ rad and then becomes approximately flat. At $r\sim500$--$1000~R_{\rm g}$, this angular scale corresponds to a transverse length scale of $\sim10$--$40~R_{\rm g}$, which is broadly consistent with $l_1\sim30~R_{\rm g}$ identified in Figure~\ref{fig:StructureFunctions}(a). We therefore interpret $l_1$ as representing the characteristic transverse width of the high-density structures. In contrast, the angular density structure functions of the non-clump regions and the entire outflow continue to increase toward larger angular separations without showing a comparably clear plateau. The characteristic transverse scale found above therefore appears to be a distinctive property of the high-density structures.

The fact that the density structure function of the clump regions remains approximately flat up to the much larger scale $l_2\sim500~R_{\rm g}$ further suggests that the high-density structures are not approximately isotropic clumps with a characteristic size of $\sim30~R_{\rm g}$, but instead have an elongated morphology, with a transverse width of several tens of $R_{\rm g}$ and structure extending to spatial scales substantially larger than this width.

The structure-function analysis indicates that the morphology is consistent with the finger-like structures expected from RT instability. On the other hand, velocity shear is also present in the outflow, so KH instability may operate as well. We therefore estimate the growth timescales of RT and KH instabilities below.

To examine the growth of RT and KH instability, we estimated their characteristic growth timescale. The RT timescale is \citep[e.g.,][]{Chandrasekhar+1962,Sharp+1984}
\begin{equation}
t_{\rm RT}=\frac{1}{\gamma_{\rm RT}}=\sqrt{\frac{1}{Akg_{\rm eff}}},
\end{equation}
with linear RT growth rate
\begin{equation}
\gamma_{\rm RT} = \sqrt{A k g_{\rm eff}},
\end{equation}
where $A$ is the density contrast
\begin{equation}
A=\frac{\rho_{\rm high}-\rho_{\rm low}}{\rho_{\rm high}+\rho_{\rm low}},
\end{equation}
$k=2\pi/\lambda$ is the wavenumber for the perturbations, and $g_{\rm eff}$ is the effective acceleration.
Adopting typical values for clumps from \citet{Hu+2026}, the typical RT timescale is
\begin{equation}
t_{\rm RT} \sim 10^4~{\rm s} \cdot \left( \frac{\lambda}{100 R_{\rm g}}\right)^{1/2} 
    \left[ \frac{A}{9/11} \cdot \frac{g_{\rm eff}}{10 \cdot g(500 R_{\rm g})}\right]^{-1/2}.
\end{equation}
with $g(500 R_{\rm g})$ the gravity at $R=500~R_{\rm g}$. The KH growth timescale is \citep[e.g.,][]{Chandrasekhar+1962}
\begin{equation}
t_{\rm KH}= \frac{1}{\gamma_{\rm KH}} = \frac{\rho_1+\rho_2} {k\,\Delta v\,\sqrt{\rho_1\rho_2}},
\end{equation}
with $\gamma_{\rm KH}$ the KH growth rate, $\Delta v$ the velocity difference between the shears, and wavelength $k=2\pi/\lambda$.
Assuming KH instability occurs with $\lambda\sim 100~R_{\rm g}$, $\rho_1=10\rho_2$, and $\Delta v\sim 0.01 c$, the timescale is 
\begin{equation}
t_{\rm KH}\sim 3\times 10^5~{\rm s}.
\end{equation}
The estimated characteristic timescale for RT instability is one order of magnitude smaller than the dynamical, while timescale for KH instability is on the same order as dynamical/orbital timescale. It suggests that there is plenty time for RT instability to develop, while time is marginal for KH instability. 

Combining the above timescale estimates with the elongated morphology independently inferred from the structure-function analysis suggests RT instability as a plausible mechanism for producing the high-density structures. However, RT instability cannot be identified as the unique formation mechanism, and a contribution from KH instability cannot be excluded. Higher-resolution simulations will be required to clarify its role (see Appendix~\ref{app:B} for quantification for KH instability).

\subsection{Effects of Disk Boundary Conditions}

Our numerical experiments demonstrate that spatial fluctuations at the disk surface play a crucial role in the development of clumpy outflows and strongly support RT instability as the underlying fragmentation mechanism. These findings also motivate a careful reassessment of our numerical setup. In the present 2D GR-RHD simulations, magnetic fields and viscosity are neglected, simplifying the physical processes governing the accretion flow. As a consequence, the accretion disk cannot be sustained self-consistently throughout the simulation. To circumvent this limitation, we prescribe disk boundary conditions based on the radiation-hydrodynamic simulations of \citet{Yoshioka+2024}. Although this treatment successfully reproduces clumpy outflows, the disk boundary is not evolved self-consistently and therefore remains an approximation. Consequently, it limits our ability to investigate the coupling between the accretion disk and the outflow, as well as the early growth of the hydrodynamic instabilities responsible for clump formation. 
In particular, given the significance of the spatial fluctuations at the disk surface in clump formation, the absence of fluctuations in the azimuthal direction in current 2D simulations may be another important limitation of the present setup. The nonlinear development and fragmentation of the RT structures may differ in three dimensions.

We expect fully self-consistent three dimensional GR-RMHD simulations, where the accretion disk is sustained through magnetic stresses and naturally develops spatial perturbations, will be essential for achieving a comprehensive understanding of the origin and evolution of clumpy outflows. In such 3D GR-RMHD simulations, magnetorotational instability (MRI) is expected to maintain disk turbulence and naturally generate the perturbations that seed RT instability at the disk surface. Such simulations will provide a self-consistent framework for investigating the interplay between disk turbulence, radiation-driven instabilities, and the formation of clumpy outflows.

\section{Conclusions} \label{sec:conclusion}

In this paper, we perform a series of general-relativistic radiation-hydrodynamic simulations to investigate the formation of clumpy outflows under different boundary-condition configurations. Through a set of controlled numerical experiments, we systematically examine the roles of the inner and disk boundary conditions by varying their spatial and temporal properties. We identify the key numerical conditions required for clump formation and perform structure-function analyses of both clump and non-clump components to investigate the underlying fragmentation mechanism. Our main conclusions are summarized as follows:

\begin{itemize}
\item Spatial smoothing and the removal of temporal variability at the inner boundary have only a minor impact on the formation of clumpy outflows, indicating that the inner boundary conditions are not the primary factor controlling clump formation.
\item In contrast, the disk boundary conditions play a dominant role. Clumpy outflows are reproduced only when sufficient spatial fluctuations are preserved along the disk surface, whereas strongly smoothed boundary conditions largely suppress clump formation.
\item By separately smoothing individual physical quantities at the disk boundary, we find that preserving the spatial fluctuations of any one of the major hydrodynamic or radiation variables is sufficient to sustain clump formation. Furthermore, retaining spatial fluctuations in a larger number of quantities leads to more numerous clumps distributed over a broader range of polar angles.
\item Thermal instability does not appear to develop significantly in the simulation. Structure-function analysis suggests that the high-density clumpy regions have elongated morphologies, possibly induced by radiation RT instability while KH instability cannot be ruled out in current simulations.
\end{itemize}

Our simulations successfully reproduce clumpy outflows with properties broadly consistent with those inferred for the ultra-fast outflow in PDS~456 \citep{Xrism+2025}. Through controlled numerical experiments, we identify spatial fluctuations along the disk surface as the key prerequisite for reproducing clumpy outflows in the present simulations. 
The elongated morphology inferred from the structure-function analysis suggests that radiation Rayleigh--Taylor instability may play a role in the formation of these structures.
Although the prescribed disk boundary conditions enable us to isolate the underlying physical processes, they prevent a fully self-consistent treatment of the coupling between the accretion disk and the outflow. Future three dimensional general-relativistic radiation magnetohydrodynamic simulations, in which disk turbulence and magnetic stresses naturally generate the seed perturbations in both radial and azimuthal directions, will be essential for establishing a complete picture of the origin and evolution of clumpy outflows and their dependence on black hole accretion properties.

\begin{acknowledgments}
The numerical calculations and simulations were performed partially on HPE Cray XD2000 at the Center for Computational Astrophysics (CfCA), National Astronomical Observatory of Japan and partially on Yukawa-21 at Yukawa Institute for Theoretical Physics (YITP) in Kyoto University.

This research was supported by the Japan Society for the Promotion of Science (JSPS) through KAKENHI Grant Numbers JP24KF0130 (HH, KO), JP21H04488 (HRT, KO), JP24K00678 (HRT, KO), JP25K01045 (YA, KO), JP24K00672 (HRT), and JP23K03445 (YA). This work was also supported by MEXT as “Program for Promoting Researches on the Supercomputer Fugaku” (Structure and Evolution of the Universe Unraveled by Fusion of Simulation and AI; Grant Number JPMXP1020240219; YA, HRT, KO), 
by Joint Institute for Computational Fundamental Science (JICFuS, KO), and
(in part) by the Multidisciplinary Cooperative Research Program in CCS, University of Tsukuba. 
This work was also supported in part by JST SPRING Grant No. JPMJSP2110 (SY).

The authors acknowledge the utilization of artificial intelligence (AI), such as ChatGPT, to assist with English-language editing and polishing of analysis scripts.
\end{acknowledgments}

\section*{Data Availability}

The data underlying this article will be shared upon reasonable request to the corresponding author.
The analysis scripts and processed data necessary to reproduce the figures presented in this work may be made available upon reasonable request.

\appendix
\section{Simulation Case without Radiation}\label{app:A}

\begin{figure}[htb]
\centering
\includegraphics[width=0.65\linewidth]{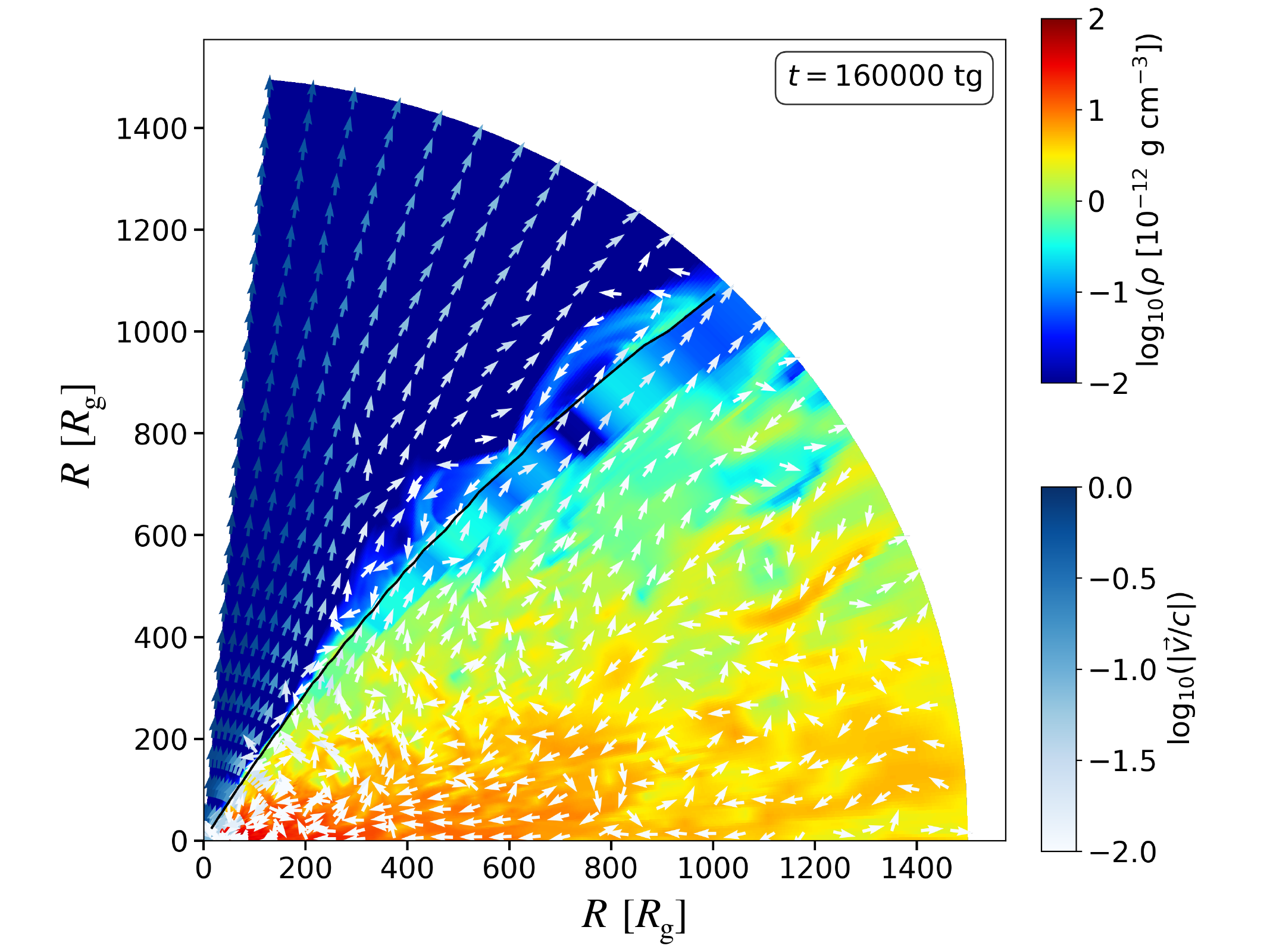}
\caption{
Two-dimensional density distribution for the simulation without radiation. The arrows indicate the velocity field. Above the disk, the outflow consists of smooth, low-density gas accelerated solely by gas pressure. In the absence of radiative support, gas near the disk surface falls back toward the black hole, and no clump-like structures are formed.
\label{fig:NoRadSim}}
\end{figure}

We present the two-dimensional density distribution for a simulation performed without the radiation module in Figure~\ref{fig:NoRadSim}. The numerical setup is identical to that of the DBC\_Seg\_30Rg model, except that radiation is excluded. In contrast to the DBC\_Seg\_30Rg case, the outflow above the disk consists of smooth, low-density gas without any clump-like structures. In this case, the outflow is driven solely by gas pressure. We also observe gas inflow near the disk surface. Without radiative support, the gas loses its outward acceleration and falls back toward the black hole. These results demonstrate that radiation is indispensable for the formation of clumpy outflows in our simulations.

\section{Vorticity for clumps and non-clumps}\label{app:B}

\begin{figure}[htb]
\centering
\includegraphics[width=0.65\linewidth]{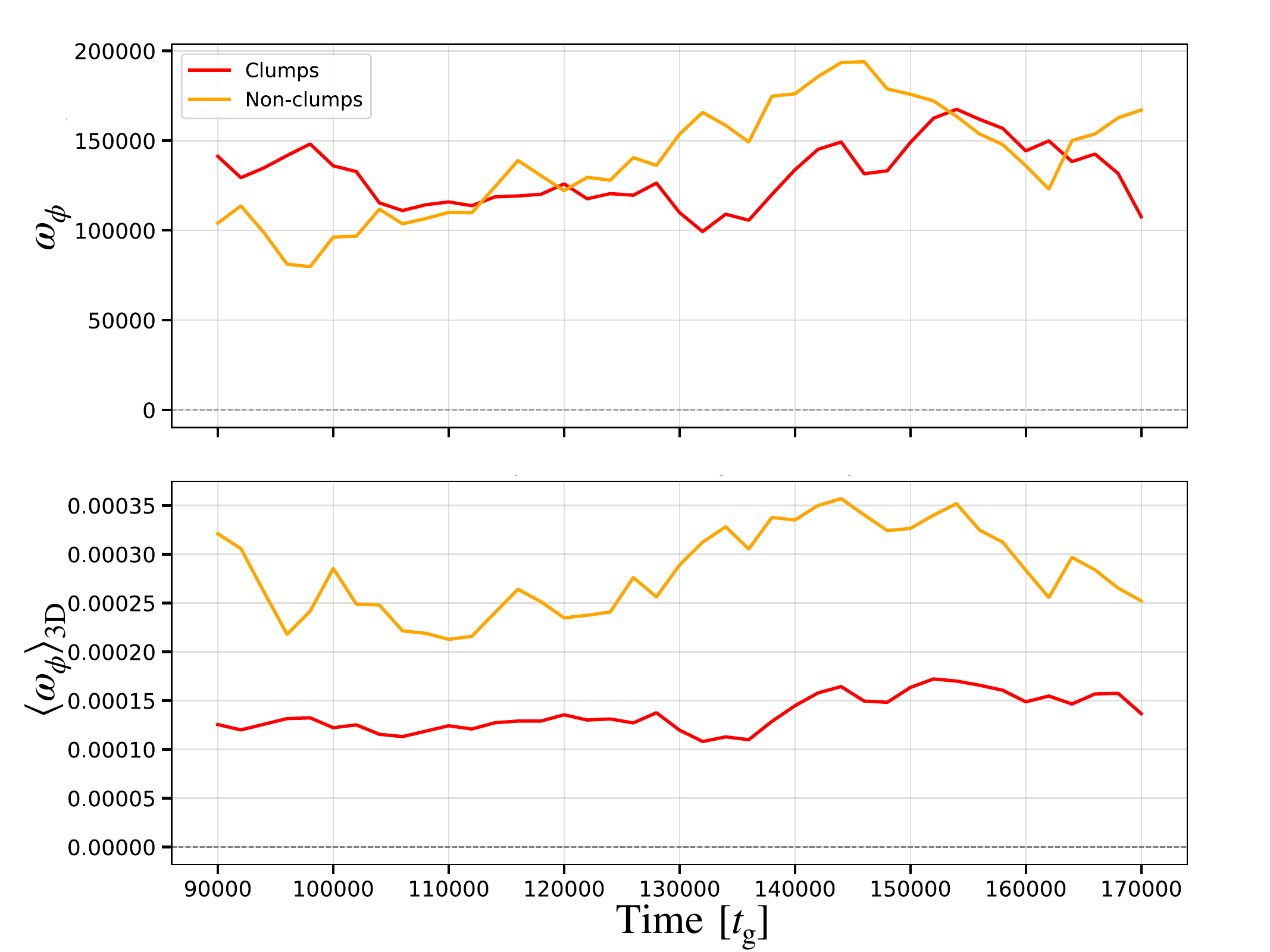}
\caption{
Time evolution of the vorticity in the Fiducial simulation. The upper panel shows the total vorticity, while the lower panel presents the volume-averaged vorticity for the clump (red) and non-clump (orange) components.
\label{fig:Vortex}}
\end{figure}

We further examine the role of KH instability by comparing the evolution of the vorticity in clumps and non-clumps. Figure~\ref{fig:Vortex} presents the time evolution of the vorticity for the Fiducial model. In the two-dimensional spherical coordinate system, the azimuthal component of the vorticity is defined as
\begin{equation}
\omega_{\phi} =\frac{1}{r} \cdot \left[  \frac{\partial(r\cdot v_{\theta})}{\partial r} - \frac{\partial v_r}{\partial \theta}\right],
\end{equation}
where $v_r$ and $v_\theta$ denote the radial and polar velocity components, respectively. 
The total vorticity is computed as
\begin{equation}
\Omega_{\phi} =2\pi \iint \omega_{\phi}  r^2 \sin\theta \cdot {\rm d}r {\rm d}\theta,
\end{equation}
whereas the volume-averaged vorticity is defined as
\begin{equation}
\langle \omega_{\phi} \rangle_{\rm 3D} =\frac{2\pi}{V} \iint \omega_{\phi}  r^2 \sin\theta \cdot {\rm d}r {\rm d}\theta,
\end{equation}
where $V$ is the volume occupied by the clump and non-clump components, respectively. During the calculation, we exclude the region with $\theta<15^\circ$ to avoid contamination from the high-velocity polar outflow.
The upper panel of Figure~\ref{fig:Vortex} shows that the total vorticities of the clump and non-clump components are nearly identical throughout the simulation. In contrast, the lower panel demonstrates that the volume-averaged vorticity within the clumps is systematically lower than that of the ambient gas by approximately a factor of two. Combined with the structure-function analysis presented in Section~\ref{sec:discussion}, these results suggest that KH instability does not play a dominant role in the formation of clumpy outflows. Alternatively, KH instability may develop only on spatial scales that are not yet resolved by the present simulations.

\newpage

\end{CJK*}
\end{document}